\documentclass{emulateapj}
\shorttitle{Spectropolarimetry of SN~2011fe}
\shortauthors{Smith et al.}
\begin{document}

\title{Multi-epoch Spectropolarimetry of SN~2011\lowercase{fe}}

\author{Paul S.\ Smith\altaffilmark{1}, G.\ Grant
  Williams\altaffilmark{1,2}, Nathan Smith\altaffilmark{1}, Peter A.\
  Milne\altaffilmark{1}, Buell T. Jannuzi\altaffilmark{3}, \& E.M.\
  Green\altaffilmark{1}}

\altaffiltext{1}{University of Arizona, Steward Observatory, 933 N.\
  Cherry Ave., Tucson, AZ 85721.}
\altaffiltext{2}{MMT Observatory, 933 N.\ Cherry Ave., Tucson, AZ
  85721.}
\altaffiltext{3}{National Optical Astronomy Observatory, 950 N.\
  Cherry Ave., Tucson, AZ 85719-4933.}

\begin{abstract}

  We present multiple spectropolarimetric observations of the nearby
  Type~Ia supernova (SN) 2011fe in M101, obtained before, during, and
  after the time of maximum apparent visual brightness.  SN~2011fe
  exhibits time-dependent polarization in both the continuum and
  strong absorption lines.  At all epochs, red wavelengths exhibit a
  degree of continuum polarizaiton of 0.2--0.4\%, likely indicative of
  persistent asymmetry in the electron-scattering photosphere.
  However, the degree of polarization across the Si~{\sc ii}
  $\lambda$6355 absorption line varies dramatically from epoch to
  epoch.  Before maximum, Si~{\sc ii} $\lambda$6355 shows enhanced
  polarization at the same position angle (PA) as the polarized
  continuum.  During two epochs near maximum, however, Si~{\sc ii}
  $\lambda$6355 absorption has a lower degree of polarization, with a
  PA that is 90\arcdeg\ from the continuum.  After maximum, the
  absorption feature has the same degree of polarization and PA as the
  adjacent continuum.  Another absorption feature in the blue (either
  Si~{\sc ii} $\lambda$5051 or a blend with Fe~{\sc ii} lines) shows
  qualitatively similar changes, although the changes are shifted in
  time to an earlier epoch.  This behavior is similar to that seen in
  broad absorption-line quasars, where the polarization in absorption
  features has been interpreted as the line absorbing some of the
  unpolarized continuum flux.  This behavior, along with the
  90\arcdeg\ shifts of the polarization PA with time, imply a
  time-dependent large-scale asymmetry in the explosion.

\end{abstract}

\keywords{supernovae: general}

\section{INTRODUCTION}

Type~Ia supernova (SN~Ia) explosions convey information about the
nucleosynthesis of the thermonuclear destruction of a CO white dwarf
(Iwamoto et al.\ 1999), and they provide a way to measure the
expansion of the universe by using their peak magnitudes as
standardizable candles (Phillips 1993). The unexpected finding that
the expansion of the universe is accelerating (Riess et al.\ 1998;
Perlmutter et al.\ 1998) has focussed interest on a better
understanding of the SN~Ia explosion mechanism.  It has long been
recognized that there are variations within the SN~Ia category. More
luminous events rise to peak and decline from peak on a longer
timescale than less luminous events (Branch, Fisher \& Nugent 1993;
Phillips 1993).  The majority of events fall within a ``normal''
grouping, although some cases have been recognized where the
luminosity does not correlate with peak width (e.g., Benetti et al.\
2005; Wang et al.\ 2009; Foley \& Kasen 2011).  Asymmetries in the
explosion may hold important clues to the explosion mechanism itself,
as well as the consequent diversity in observed properties.

Spectropolarimetry has emerged as a powerful probe of SNe~Ia (see
Livio \& Pringle 2011).  The degree of polarization of the continuum
emission is generally lower for SNe~Ia than for core-collapse events,
but it has been detected at significant levels for a range of SN~Ia
sub-classes.  Measured polarization at the wavelengths of observed
absorption lines is particularly interesting, as it affords the
opportunity to study the distribution of specific elements within the
ejecta. This ``line polarization'' has been observed to change
markedly near maximum light.  The signature Si~{\sc ii}
$\lambda$6355\AA\ line has exhibited line polarization in a number of
SNe~Ia: SNe~1999by (Howell et al.\ 2001), 2001el (Wang et al.\ 2003),
2002bf, 1997dt, 2003du, 2004dt (Leonard et al.\ 2005), 2004dt (Wang et
al.\ 2006), and 2006X (Patat et al.\ 2009).  Absorption lines of other
elements have also been detected, notably Fe~{\sc ii} lines
(SN~1997dt: Leonard et al.\ 2005), and the Ca~{\sc ii} IR triplet
(SN~2001el: Wang et al.\ 2003; SN~2006X: Patat et al.\ 2009).  For a
review of polarimetric studies of SNe~Ia, see Wang \& Wheeler (2008).

SN~2011fe occurred in M101, and was discovered on 2011 August 24 by
the Palomar Transient Factory (PTF: Brown et al.\ 2011, Nugent et al.\
2011a, 2011b). The proximity of M101, $\sim$6.2 Mpc, and the location
of the SN far from the host galaxy core and spiral arms, suggested
that the SN would become the brightest SN~Ia since SN~1972E. It was
predicted to have a visual peak brighter than 10.0 mag. Studies of the
light curve suggest that the SN was discovered just 0.5 days after the
explosion, and the explosion time is constrained to very high
precision (Nugent et al.\ 2011b). Optical spectra revealed SN~2011fe
to be a normal SN~Ia, with detections of C~{\sc ii} $\lambda$6580 and
$\lambda$7234 in absorption (Cenko et al.\ 2011).  Studies of
pre-explosion images of the site of SN~2011fe place the strictest
upper limits yet on the luminosity of any SN~Ia progenitor, arguing
against a single-degenerate progenitor containing a giant donor star
(Li et al.\ 2011).

SN~2011fe was the nearest Type~Ia explosion in several decades,
providing an unprecedented opportunity to obtain spectropolarimetry of
a normal SN~Ia with modest-aperture telescopes. We initiated a
campaign to obtain multi-epoch spectropolarimetry of SN~2011fe at
Steward Observatory ({\it SO}), using the 1.54-m Kuiper and 2.3-m Bok
telescopes.  We describe the results of these observations below.

\begin{figure*}
\epsscale{0.7} 
\plotone{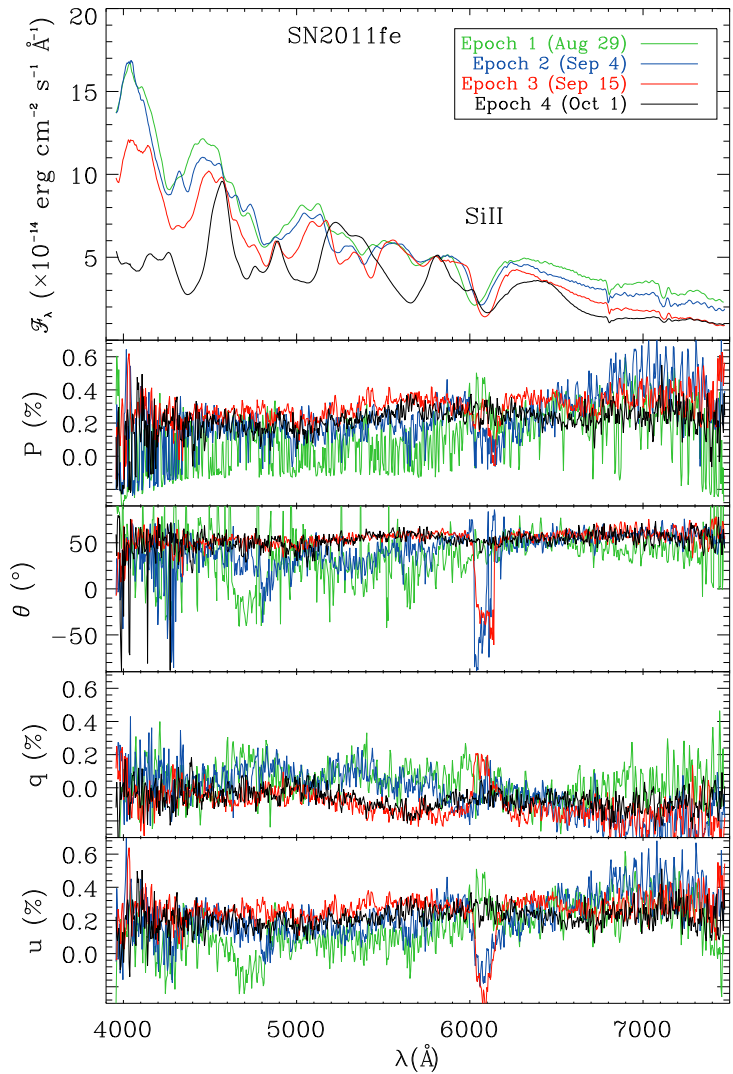}
\caption{The four epochs (color coded) of flux, degree of
  polarization, angle, and normalized Stokes parameters q and u.  The
  flux values have been scaled to the first epoch flux at 5800 \AA \
  using the following flux ratios: 1.0, 0.228, 0.148, 0.215.  For the
  first epoch, angles less than $-$45\arcdeg\ have been rotated by
  180\arcdeg\ to remove confusion near the feature at 4700 \AA.}
\label{fig:spec-seq}
\end{figure*}

\begin{table}\begin{center}\begin{minipage}{3.25in}
      \caption{Observation log for SN~2011\lowercase{fe}}
\scriptsize
\tighten
\begin{tabular}{@{}lcccccc}\hline\hline
UT Date &days &days & UT &UT & Exp  &Epoch \\
             &+exp\tablenotemark{a} &  $+$B$_{\rm
               max}$\tablenotemark{b} & (start) & (end) & (sec) & \\
\hline
2011-08-29       & 5.4  & -12.0 & 03:32:02  &  04:47:55  & 3840  &1      \\
2011-09-04\tablenotemark{c} & 11.4 & -6.0 & 02:51:44  &  04:33:19  & 4320  &2 \\
2011-09-15       & 22.4 &   5.0 & 02:53:41  &  03:40:45  & 1504  &3      \\
2011-09-16       & 23.4 &   6.0 & 02:48:08  &  03:58:29  & 2784  &3      \\   
2011-09-26       & 33.4 & 16.0 & 02:41:28  &  03:14:17  & 1440  &4      \\
2011-09-28       & 35.4 & 18.0 & 02:53:44  &  03:13:46  & 800   &4      \\ 
2011-09-29       & 36.4 & 19.0 & 02:20:22  &  03:06:02  & 1920  &4      \\
2011-09-30       & 37.4 & 20.0 & 02:26:08  &  02:56:33  & 1920  &4      \\ 
2011-10-01       & 38.4 & 21.0 & 02:24:33  &  02:54:56  & 1280  &4      \\ 
2011-10-06       & 43.4 & 26.0 & 02:09:04  &  02:39:28  & 1280  &4      \\
\hline
\end{tabular}
\tablenotetext{a}{Epoch relative to explosion, 23.7 Aug.\ 2011, as
  estimated by averaging explosion dates of Brown et al.\ (2011) and
  Nugent et al.\ (2011).}  
\tablenotetext{b}{Epoch relative to the reported date of B$_{\rm
    max}$, 10.1 $+/-0.2$ Sep. 2011, as reported by Matheson et al.\
  (2011).}  
\tablenotetext{c}{This epoch of observations was obtained with SPOL
  mounted on the 1.54m Kuiper telescope at Mt.\ Bigelow, AZ.  All
  other epochs used SPOL mounted on the 2.3-m Bok telescpe on Kitt
  Peak, AZ.}
\label{tab:obslog}
\end{minipage}\end{center}
\end{table}

\begin{figure*}
\epsscale{0.7} 
\plotone{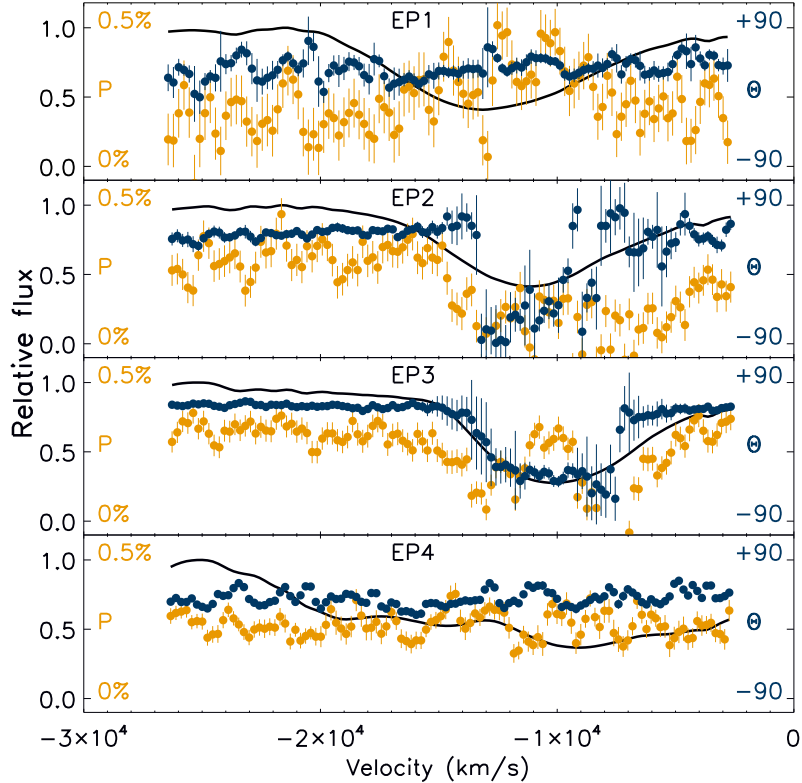}
\caption{Plots of Si~{\sc ii} $\lambda$6355 in velocity space for all
  four epochs. The flux is shown in black, $P$ in orange and $\theta$
  in blue.  Flux spectra have been normalized to highest flux in
  wavelength range.}
\label{fig:si2-vel}
\end{figure*}


\section{OBSERVATIONS}

The SPOL CCD Imaging/Spectropolarimeter \citep{schmidt92a} mounted on
the Steward Observatory 2.3-m Bok telescope (Kitt Peak, AZ) and the
1.54 Kuiper telescope (Mt.\ Bigelow, AZ) was used to obtain
spectropolarimetry of SN~2011fe over 10 nights. We have grouped the 10
nights of observations into four Epochs in Table~1.  Observations
covered 4000--7550\AA\ at a resolution of $\sim$20~\AA\ (600 line
mm$^{-1}$ grating in first order, using a
5$\farcs$1$\times$51$\arcsec$ \ slit and a Hoya L38 blocking filter).
A rotatable semiachromatic half-wave plate was used to modulate
incident polarization and a Wollaston prism in the collimated beam
separated the orthogonally polarized spectra onto a thinned,
anti-reflection-coated 800$\times$1200 SITe CCD.  The efficiency of
the wave plate as a function of wavelength is measured by inserting a
fully-polarizing Nicol prism into the beam above the slit.  A series
of four separate exposures that sample 16 orientations of the wave
plate yields two independent, background-subtracted measures of each
of the normalized linear Stokes parameters, $q\/$ and $u\/$.  Each
night, several such sequences of observations of SN~2011fe were
obtained and combined, with the weighting of the individual
measurements based on photon statistics. The polarization results for
September 15 and 16 were indistinguishable, so they were combined to
yield the final result for the third observational epoch.  Similarly,
the polarization spectra from the six observations obtained between
September 26 and October 6 (Epoch 4) were averaged together, since we
detected no inter-night variations in $Q$ or $U$ over this time
period.

We confirmed that the instrumental polarization of SPOL mounted on the
Bok and Kuiper telescopes is much less than 0.1\% through observations
of the unpolarized standard stars BD+28$^{\circ}$4211 and HD~212311
\citep{schmidt92b} during each epoch.  The linear polarization
position angle on the sky ($\theta\/$) was determined by observing the
interstellar polarization standards Hiltner~960 and VI~Cyg~\#12
\citep{schmidt92b} during all epochs.  Additional observations of the
polarization standard stars BD+59$^{\circ}$389 and BD+64$^{\circ}$106
were made during the third epoch (Table~1). The adopted correction
from the instrumental to the standard equatorial frame for $\theta\/$
for all epochs was determined from the average position angle offset
of Hiltner~960 and VI~Cyg~\#12. Differences between the measured and
expected polarization position angles were $< 0\farcs3$ for all of the
standard stars.

During the first epoch, two field stars within $\sim$2\arcmin \ of
SN~2011fe (2MASS J14031367+5415431 and 2MASS J14025413+5416288) were
measured to check for significant Galactic interstellar polarization
(ISP) along the line-of-sight to the SN.  These stars yielded a
consistent estimate for Galactic ISP, with $P_{max} = 0.11 \pm 0.03$\%
at $\theta = 114^{\circ} \pm 7^{\circ}$ for 2MASS J14031367+5415431
and $P_{max} = 0.16 \pm 0.03$\% at $\theta = 109^{\circ} \pm
6^{\circ}$ for 2MASS J14025413+5416288, assuming that $\lambda_{max}$,
the wavelength where the interstellar polarization is at a maximum
($P_{max}$) is 5550\AA.  The results for the field stars were averaged
and $P_{max} = 0.13$\% at $\theta = 112^{\circ}$ was adopted as the
Galactic ISP in the sightline to SN~2011fe.  This low value for the
Galactic ISP is consistent with the high Galactic latitude of M101 and
the very low estimated amount of extinction for the supernova.  The
polarization spectra of SN~2011fe have been corrected for this level
of Galactic ISP assuming that it is fit well by a Serkowski law
\citep{wilking80,serkowski}.  No estimate or correction for ISP within
M101 at the location of SN~2011fe has been made (although see \S3).
Our reported values for the degree of linear polarization, $P\/$, have
been corrected for statistical bias \citep{wardle74}.

\section{POLARIZATION OF SN~2011\lowercase{fe}}

Our sequence of spectra are shown in the top panel of
Figure~\ref{fig:spec-seq}, displaying the emergence of absorption
features typical of SNe~Ia.  The continuum emission is polarized with
the red wavelengths more highly polarized than the blue wavelengths at
early epochs, reaching up to $\sim$0.4\%. The polarization of the red
continuum (6500-7500\AA) exhibits a slight decrease with time, from
about 0.4\% down to 0.2\%, while continuum polarization in the
5000-5900\AA\ range increases from undetected up to 0.2\%.

The polarization of absorption lines is clearly present in blueshifted
Si~{\sc ii} $\lambda$6355\AA\ absorption, and it changes markedly with
time.  This line polarization of Si~{\sc ii} is shown in velocity
space in Figure~\ref{fig:si2-vel}.  Before maximum at Epoch 1, Si~{\sc
  ii} $\lambda$6355\AA\ shows polarization at the same position angle
(PA) as the continuum, but is roughly 0.2\% stronger than the adjacent
continuum in polarization degree.  In the subsequent two epochs near
maximum (Epochs 2 and 3), however, Si~{\sc ii} $\lambda$6355\AA\
absorption has $\sim$0.2\% {\it weaker} polarization than the
continuum, and the absorption-line PA changes by about 90\arcdeg\
(Figure~\ref{fig:si2-vel}).  Well after maximum in Epoch 4, the
absorption trough returns to showing the same polarization and PA as
the adjacent continuum --- although note that the continuum
polarization has increased since Epoch 1.  The blueshift of the
Si~{\sc ii} feature decreases with epoch, consistent with a low
velocity gradient event (LVG; Benetti et al.\ 2005).  


Line polarization is also present in an absorption feature in the
4600-5000 \AA\ region. This might be Si~{\sc ii} $\lambda$5051, which
may be blended with nearby Fe~{\sc ii} lines.  Similar blue features
have been seen to exhibit line polarization in other SNe~Ia (Leonard
et al.\ 2005), although not necessarily the same absorption feature as
in SN~2011fe.  The polarization of this feature is also variable,
exhibiting overall behavior similar to that of Si~{\sc ii}
$\lambda$6355, but shifted about one epoch earlier in time.  In other
words, this blue absorption feature shows line depolarization in Epoch
1, which then disappears in Epochs 2 and 3 when the same
depolarization appears in the Si~{\sc ii} $\lambda$6355\AA\ line.
Moreover, the apparent depolarization of the blue line approaches a PA
that is nearly orthogonal to the continuum polarization PA, similar to
what occurs later for the red absorption line.  We speculate that the
blue absorption feature traces the same asymmetric layers in the SN
atmosphere that are seen in Si~{\sc ii} $\lambda$6355 in Epochs 2 and
3, but at an earlier time.  This may be an important clue to the
changing structure in the receding SN photosphere, which can hopefully
provide constraints on additional radiative transfer models of the
observed polarization behavior.

\begin{figure}
\epsscale{1.5} 
\plotone{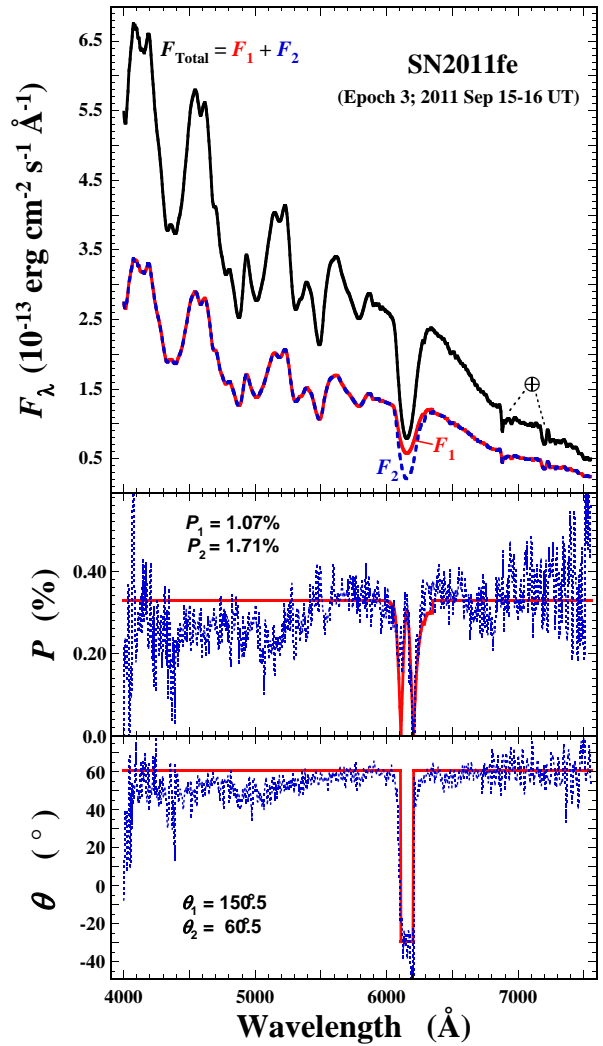}
\caption{A simple non-unique model that can match the observed
  polarization at Epoch 3 using two scattering components with
  orthogonal PA.  The two components have slightly different continuum
  polarization, and also have slightly different absorption equivalent
  widths in Si~{\sc ii} $\lambda$6355 that result in different
  residual amounts of polarization in and out of the absorption line.
  The strong shift in PA occurs when one component dominates the
  polarization PA in the absorption line.}
\label{fig:model}
\end{figure}

\section{SUMMARY AND INTERPRETATION}

A key result of our spectropolarimetry is that Si~{\sc ii}
$\lambda$6355 absorption exhibits a polarization angle that changes
with time, and more importantly, reaches a PA in the absorption line
polarization that is roughly {\it orthogonal} to the adjacent
continuum polarization angle for an extended time around maximum light
when the line depolarization is strongest.  This behavior can be
approximated by two polarized components that have perpendicular PAs,
and comparable amounts of polarization that do not completely cancel,
leaving a net residual linear polarization.  Small changes in the
strength of the absorption line feature's polarization can shift the
residual net polarization by 90\arcdeg, because one or the other
components dominates.  This behavior is approximated by a simple
two-component model of the spectrum at Epoch 3, shown in
Figure~\ref{fig:model}, where the observed flux is composed of two
such orthogonal components, but where one of the two has a different
absorption equivalent width in Si~{\sc ii} $\lambda$6355.  There is a
small amount ($\sim$0.2\%) of net residual continuum polarization
dominated by one component, but this changes across the line feature
due to absorption of the polarized continuum flux by the line.

The fact that these two components have orthogonal PAs that persist
with time and are seen at the earliest epoch in a different absorption
feature, is suggestive that there is some small amount of global
asymmetry in the ejecta of SN~2011fe, perhaps even suggesting axial
symmetry in the event.  For example, the observations could be
explained if the continuum polarization arises from an electron
scattering photosphere that is slightly elongated in the polar
direction, whereas the strongest Si~{\sc ii} $\lambda$6355 absorption
may occur in an equatorial belt.  This is just one possible scenario.
More definitive constraints on the SN ejecta geometry will require
more detailed analysis with radiative transfer calculations.

\acknowledgements
\footnotesize

P.S.S.\ acknowledges support from NASA/{\it Fermi} Guest Investigator
grant NNX09AU10G.  P.A.M.\ acknowledges support from NASA ADP grant
NNX10AD58G. BTJ acknowledges support from the NSF, through its funding
of NOAO, which is operated by AURA, Inc., under a cooperative
agreement with the NSF.

\end{document}